\documentclass[twocolumn,prb,superscriptaddress,noshowpacs, aps, 10pt]{revtex4-2}
 
\usepackage{graphicx,nicefrac}
\usepackage{amssymb} 
\usepackage{amsfonts}
\usepackage{xcolor}
\usepackage{amsmath}
\usepackage{siunitx}
\usepackage{soul} 

\usepackage[utf8]{inputenc}

\usepackage{hyperref}

\usepackage{amsmath}
\usepackage{mathtools}
\usepackage{braket}
\usepackage{dsfont}
\usepackage{bigints}
\usepackage{graphicx}
\usepackage{amssymb}
\usepackage{xcolor}
\hypersetup{
	colorlinks,
	linkcolor={red!95!black},
	citecolor={green!60!black},
	urlcolor={blue!95!black}
}

\usepackage{bbold}

\usepackage{soul}

\graphicspath{ {./Images/} }

\newcommand{\approxpropto}{\mathrel{\vcenter{
			\offinterlineskip\halign{\hfil$##$\cr
				\propto\cr\noalign{\kern2pt}\sim\cr\noalign{\kern-2pt}}}}}

\usepackage{adjustbox}
\makeatletter
\newcommand{\fitimage}[2][\@nil]{
	\begin{figure}
		\begin{adjustbox}{width=0.9\textwidth, totalheight=\textheight-2\baselineskip-2\baselineskip,keepaspectratio}
			\includegraphics{#2}
		\end{adjustbox}
		\def\tmp{#1}%
		\ifx\tmp\@nnil
		\else
		\caption{#1}
		\fi
	\end{figure} 
}

\begin{document}

\title{Laughlin's quasielectron as a non-local composite fermion}
\author{Alberto Nardin}
\email{alberto.nardin@universite-paris-saclay.fr}
\affiliation{Pitaevskii BEC Center, INO-CNR and Dipartimento di Fisica, Universit\`a di Trento, I-38123 Trento, Italy}
\affiliation{Universit\'e Paris-Saclay, CNRS, LPTMS, 91405 Orsay, France}

\author{Leonardo Mazza}
\email{leonardo.mazza@universite-paris-saclay.fr}
\affiliation{Universit\'e Paris-Saclay, CNRS, LPTMS, 91405 Orsay, France}

\begin{abstract}
We discuss the link between the quasielectron wavefunctions proposed by Laughlin and by Jain and show both analytically and numerically that Laughlin's quasielectron is a non-local composite fermion state.
Composite-fermion states are typically discussed in terms of the composite-fermion Landau levels (also known as Lambda levels).
In standard composite-fermion quasielectron wavefunctions the excited Lambda levels have sub-extensive occupation numbers. 
However, once the Laughlin's quasielectron is reformulated as a composite fermion, an overall logarithmic occupation of the first Lambda level is made apparent, which includes orbitals that are localized at the boundary of the droplet.
Even though the wavefunction proposed by Laughlin features a localised quasielectron with well-defined fractional charge, it exhibits some non-trivial boundary properties which motivate our interpretation of Laughlin's quasielectron as a non-local object.
This has an important physical consequence: Laughlin's quasielectron fractionalizes an incorrect spin, deeply related to the anyonic braiding statistics.
We conclude that Laughlin's quasielectron is not a good candidate for a quasielectron wavefunction.
\end{abstract}

\maketitle

\textit{\textbf{Introduction.}}
In a seminal paper on the theory of the fractional quantum Hall effect (FQHE), Laughlin introduced his celebrated wavefunctions for describing the bulk of a FQHE electron liquid at filling $\nu = 1/m$, with $m$ an odd integer~\cite{Laughlin_1983}. 
The same article also describes model wavefunctions for fractionally-charged excitations in the liquid's bulk, dubbed quasiholes (QHs) and quasielectrons (QEs)~\cite{Laughlin_1983}.
Whereas the wavefunctions for the bulk and for the QH are the starting point of our understanding of the FQHE, the QE wavefunction has experienced a more controversial history.

Even though a theoretical analysis~\cite{WilczekShapere_GeometricPhases_1989} similar to that applied to the QH~\cite{Arovas_1984} suggests that Laughlin's QE (LQE) has the proper charge and braiding statistics, some inconsistencies were soon pointed out.
At the end of the '90s, a series of numerical works showed that LQE has the correct fractional charge $-e/m$ but not the correct braiding properties~\cite{KjonsberLeinaas_IntJModPhysA_1997, KJONSBERG1999705,KJONSBERG_1999}.
The problem was subsequently revisited and it was shown that the braiding phase is plagued by $\mathcal{O}(1/N)$ effects~\cite{Jeon_2010}.
Recently, it was shown that it is possible to associate a fractional spin~\cite{Wilczek_PRL_1982a, Wilczek_PRL_1982b,Preskill_2004} to all FQHE quasiparticles, not only on the sphere~\cite{Li_1992,Li_1993,Einarsson_1995,Read_Arxiv2008_Spin,Gromov_2016,Trung_PRB_2023}, but on the plane as well~\cite{Macaluso_PRR_2020,Comparin_2022,Nardin_PRB_2023}: 
LQE has a spin that does not coincide with that of the anti-anyon of Laughlin's QH~\cite{Nardin_PRB_2023}.
All of these problems hindered its usefulness and made it unsuitable for the interpretation of experiments and numerical simulations. 
In general, identifying the deep reason causing the flaws of a trial wavefunction is a way to progress in our understanding of the FQHE.
Moreover, this understanding is even more important here because the elegance and simplicity of Laughlin's construction, as well as its historical importance, still grant it a place in several FQHE lecture notes~\cite{goerbig2009quantum,tongLectureNotes,simon2016topological}.

While the aforementioned studies were published, Jain's composite fermion (CF) theory~\cite{Jain_1989,jainCompositeFermionsBook_2007} emerged as a new paradigm for studying FQHE states.
Not only this theory allows for a qualitative understanding of FQHE states in terms of an integer quantum Hall effect of CFs,
but also provides a large class of wavefunctions for FQHE ground states and excitations. 
Among these, a new CF QE wavefunction~\cite{Jeon_2003,Jeon_2004, Hansson_PRB_2007,Kjall_2018} (generalised also to non-Abelian states in Ref.~\cite{Hansson_PRL_2009}) was proposed and shown not only to have the correct charge but also the correct braiding properties~\cite{KJONSBERG1999705},
and soon emerged as the main paradigm for discussing QE excitations.
Exact diagonalization studies indeed showed that Jain's QE (JQE) has a better overlap with the realistic Coulomb QE state than LQE for small systems~\cite{Dev_PRB_1992} and has lower variational energy for larger ones~\cite{Jeon_2003b}.
It has been argued that this happens because LQE does not satisfy certain clustering conditions~\cite{Bernevig_PRL_2009}.
Note that more recently other candidate QE wavefunctions for Abelian and non-Abelian states have been proposed, which will however not be discussed here~\cite{Bochniak_CommPhys_2022, bochniak2023fusion}.

Laughlin's and Jain's approaches to the QE are very different and it is difficult to establish a comparison between the two, and even to try to cure the difficulties of LQE in the light of Jain's theory.

In this letter we investigate and unveil the link between LQE and JQE. 
In particular, we analytically show that LQE can be rewritten as a CF QE with a long tail in the occupation of the first Lambda level (namely, the CF Landau levels) that extends all the way through the system's boundary.
We are able to introduce a class of wavefunctions which interpolate between JQE and LQE by means of a suitable truncation of the  tail.
When this latter does not reach the compressible edge, the projection of the state onto the lowest Landau level (LLL) yields a perfectly well-behaved QE that can be identified with an excited version of Jain's CF QE.
On the other hand, if such a tail does indeed extend to the boundary, we obtain a non-local object consisting of a fractional charge in the bulk and an excitation of the cloud's edge.
This non-locality is responsible for reproducing the physics of the LQE (namely, the correct fractional charge but the wrong spin)
and thus, according to our understanding, it explains all the aforementioned problems.
Finally, we compute the spin of the LQE and show that it depends on the LLL projection scheme; based on this observation we argue that LQE is not topologically robust.
For these reasons, we conclude that the LQE is not a good candidate for a QE wavefunction.

This letter is organised as follows. 
We begin by recalling the definition of LQE and JQE. We subsequently analyze the wavefunction proposed by Laughlin's within the framework of CFs and show analytically that it is indeed a non-local CF wavefunction.
We then substantiate our statements with a numerical analysis of the wavefunctions obtained by truncating in different ways the long tail of LQE. 
We finally draw our conclusions.

\textit{\textbf{QE wavefunctions.}}
We consider a 2D system of interacting quantum particles with mass $M$ and charge $q$, subject to a strong orthogonal magnetic field $\boldsymbol{B}$. 
The single-particle energy states organize in highly degenerate Landau levels
whose partial filling leads to a massive ground-state (GS) degeneracy. 
This is lifted by strong enough interparticle interactions, leading to the formation of fascinating strongly-correlated states of matter -- the so-called FQHE.
Laughlin's wavefunction~\cite{Laughlin_1983} is the most notable model wavefunction for the FQHE, and it describes the system when the LLL is fractionally filled at $\nu=1/m$, with $m$ (in the fermionic case) an odd integer:
\begin{equation}
	\Psi_L(\{ z_i \}) \sim \prod_{i<j}(z_i-z_j)^{m} \exp\left(-\frac{1}{4}\sum_i |z_i|^2\right).
\end{equation}
Here the particle coordinates are described in terms of holomorphic variables $z=(x+iy)/l_B$, as suitable for the description of the LLL in the symmetric gauge $\mathbf{A}=\frac{B}{2}(-y,x,0)$, and $l_B=\sqrt{\hbar/qB}$ is the system's magnetic length.
From now on we will drop the omnipresent Gaussian factors.

In the same article, Laughlin 
noticed that such a state can host emergent
fractionally-charged excitations that are localized in the bulk of the FQH liquid~\cite{Laughlin_1983}.
In this letter our interest will be focused on the QE
\begin{equation}
	\label{eq:LaughlinQE}
	\Psi_{LQE}(\{ z_i \}) \sim \left(\prod_i 2\frac{\partial}{\partial z_i}\right)\prod_{i<j}(z_i-z_j)^{m},
\end{equation}
that owes its name to the fact that it describes a fractional excess of charge (given by $1/m$ when measured in units of the FQHE constituents charge, $q$) that, in the case of Eq.~\eqref{eq:LaughlinQE}, is exactly placed at the centre of the system.

As we already mentioned, a different QE wavefunction based on the CF approach to the FQHE was proposed by Jain
\begin{equation}
	\label{eq:JainQE}
	\Psi_{JQE} (\{ z_i \}) \sim \hat{P}_\text{LLL} 
	\left|
	{
		\begin{array}{cccc}
			z_0^* & z_1^* & z_2^* & \hdots \\
			1 & 1 & 1 & \hdots \\
			z_0 & z_1 & z_2 & \hdots \\
			\vdots & \vdots & \vdots & \ddots \\
		\end{array}
	}
	\right|
	\prod_{i<j}(z_i-z_j)^{m-1}
\end{equation}
in an attempt to better describe the physics of the QE~\cite{Dev_PRB_1992,Jeon_2003b}. Here, the QE is placed at the system's centre and
$\hat{P}_{LLL}$ reminds us of the need of projecting the CF wavefunction to the LLL to get rid of the antiholomorphic contributions $\propto z^*$.
Indeed, energetic considerations based on the fact that the cyclotron energy $\hbar \omega_c=\hbar\,qB/M$ is much larger than the typical interaction energy suggest to neglect any population of excited Landau levels.

\textit{\textbf{Laughlin's QE in the light of the composite fermion theory.}}
Our starting point is the LQE in Eq.~\eqref{eq:LaughlinQE} in its unprojected form:
\begin{equation}
 	\Psi_{LQE} (\{ z_i \}) \sim 
 	\left(\prod_i z_i^*\right)\prod_{i<j}(z_i-z_j)^{m},
 	\label{Eq:LQE:Unprojected}
\end{equation}
from which Eq.~\eqref{eq:LaughlinQE} is obtained by applying the Girvin-Jach projection prescription~\cite{GirvinJach_PRB_1984}.
To establish a link with CF theory, we rewrite Eq.~\eqref{Eq:LQE:Unprojected} as
\begin{equation}
	\Psi_{LQE} (\{ z_i \})
	\sim \underbrace{\left(\prod_i z_i^*\prod_{i<j}(z_i-z_j)\right)}_{\Phi(\{ z_i \})}\prod_{i<j}(z_i-z_j)^{m-1}
\end{equation}
where
we separated a ``vortex attachment" term $\prod_{i<j}(z_i-z_j)^{m-1}$ 
from a non-interacting fermionic wavefunction $\Phi$ which we can conveniently write as an $N$-particle Slater determinant
\begin{equation}
	\label{eq:CF1}
	\Phi (\{ z_i \})
	=
	\left|
	{
		\begin{array}{cccc}
			z_0^* & z_1^* & z_2^* & \hdots\\
			z_0^* z_0 & z_1^* z_1 & z_2^* z_2 & \hdots \\
			z_0^* z_0^2 & z_1^* z_1^2 & z_2^* z_2^2 & \hdots \\
			\vdots & \vdots & \vdots & \ddots \\
		\end{array}
	}
	\right|.
\end{equation}

The wavefunction of each non-interacting orbital is proportional to $z^*z^l$, which has well defined angular momentum $l-1$.
These orbitals are not energy eigenstates of the single particle Landau Hamiltonian though; instead, they are coherent superpositions of the two states with angular momentum $l-1$ belonging to the lowest ($n=0$) and the first ($n=1$) Lambda levels:
\begin{equation}
\label{Eq:orbitals}
	\begin{cases}
		\phi_{n=0,{l-1}}(z) = \frac{1}{\sqrt{2\pi {(l-1)}!}} \left(\frac{z}{\sqrt{2}}\right)^{l-1}
		\\
		\phi_{n=1,l}(z) = \frac{1}{\sqrt{2\pi l!}}\left(\frac{z}{\sqrt{2}}\right)^{l-1}\left(\frac{z}{\sqrt{2}}\frac{z^*}{\sqrt{2}}-l\right).
	\end{cases}
\end{equation} 
We thus introduce the single-particle orbital
\begin{equation}
	\label{eq:normalizedState}
	\psi_{l-1}(z) \equiv \sqrt{\frac{1}{l+1}}\,\, \phi_{n=1,l}(z)+\sqrt{\frac{l}{l+1}}\,\,\phi_{n=0,l-1}(z)
\end{equation}
which carries $l-1$ units of angular momentum.
Since in Eq.~\eqref{Eq:orbitals} we have used normalized states, the orbital in~\eqref{eq:normalizedState}
is also normalized. 
More importantly, it is easy to see that $\psi_{l-1} \propto z^*\,z^{l}$, which are precisely the terms that appear in the non-interacting fermionic wavefunction Eq.~\eqref{eq:CF1}.
We therefore rewrite Eq.~\eqref{eq:CF1} in an equivalent manner (the only difference being an overall irrelevant normalization factor) as
\begin{equation}
	\label{eq:LaughlinCF}
	\Phi (\{ z_i \}) =
	\left|
	{
		\begin{array}{cccc}
			\psi_{-1}(z_0) & \psi_{-1}(z_1) & \psi_{-1}(z_2) & \hdots\\
			\psi_0(z_0) & \psi_0(z_1) & \psi_0(z_2) & \hdots \\
			\psi_1(z_0) & \psi_1(z_1) & \psi_1(z_2) & \hdots \\
			\vdots & \vdots & \vdots & \ddots \\
		\end{array}
	}
	\right|. 
\end{equation}

\begin{figure}[t]
	\begin{adjustbox}{width=0.5\textwidth, totalheight=\textheight-2\baselineskip,keepaspectratio,center}
		\includegraphics[]{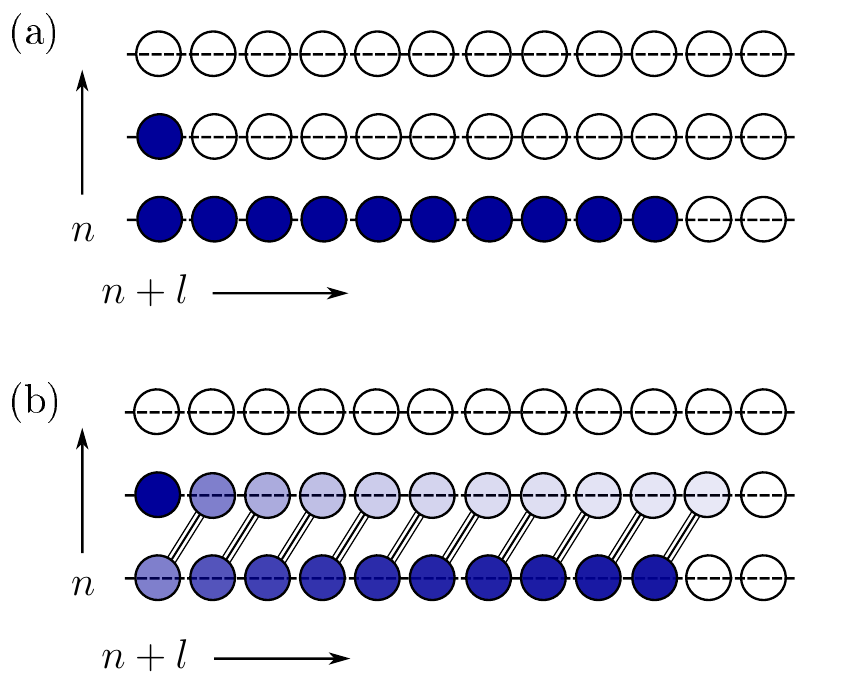}
	\end{adjustbox}
	\vspace{-0.3cm}\caption{Schematic view of (a) JQE [Eq.~\eqref{eq:JainQE}] and (b) LQE [Eq.~\eqref{eq:LaughlinQE}], in terms of their CF descriptions.
	CF states are denoted by the circles (the fluxes are not shown for graphical convenience) and labelled by their Lambda-level (index $n$) and angular momentum (index $l$).
	Empty circles denote free states, blue-filled ones the occupied ones. In (b) the transparency of the circles has been regulated according to the occupation probabilities Eq.~\eqref{eq:occupationProbabilities}.
	Diagonal bars serve as a reminder that each CF is in a superposition state of a $n=0$ and a $n=1$ Lambda-level state.
	\label{fig:QE_sketch}}
\end{figure}

The probabilities for a fermion labelled by the angular momentum $l$
to sit in the lowest or in the first excited Lambda level therefore read
\begin{equation}
	\label{eq:occupationProbabilities}
	\begin{cases}
		P_0(l) = \frac{l+1}{l+2}
		\\[3pt]
		P_1(l) = \frac{1}{l+2},
	\end{cases}
\end{equation}
which manifestly shows the long-range nature of LQE: although $\lim_{l \to \infty} P_1(l) = 0$, the decay is algebraic and interestingly the total occupation of the first Lambda level is logarithmic and diverges in the $N \to \infty$ limit. 
It would be interesting to understand whether an algebraic decay that is faster than $\sim1/l$ would lead to well-defined QE excitations.
As compared to the LQE it is easy to realize that JQE~\cite{Jeon_2003b,jainCompositeFermionsBook_2007} [Eq.~\eqref{eq:JainQE}] has basically no tail and that the first excited Lambda level is always occupied by a single fermion. 
In Fig.~\ref{fig:QE_sketch} we present a sketch comparing the two situations.

To better clarify the meaning of Eq.~\eqref{eq:occupationProbabilities},
let us consider a Laughlin state with $N$ particles: the largest occupied single-particle orbital has angular momentum $l = m(N-1)$ and is physically located at the edge of the FQHE droplet.
When a JQE is added on top of the ground state, the lowest Lambda level largest occupation rigidly shifts inwards to $m(N-1)-1$, due to a single CF being promoted to the first Lambda level.
On the other hand, if we consider a LQE, 
there is a non-negligible probability $\approx 1/m N$ to find the boundary promoted to the first Lambda level. 
We will argue in the next section that this slow algebraic decay
does indeed affect the system’s boundary after LLL projection: 
LQE has topological properties that differ from those of JQE, and in particular a different spin. 
In other words, the insertion of the LQE has non-trivial non-local effects on the system's boundary, and motivates us to interpret it as a \textit{non-local composite fermion}.

\begin{figure}[t]
	\begin{adjustbox}{width=0.5\textwidth, totalheight=\textheight-2\baselineskip,keepaspectratio,center}
		\includegraphics[]{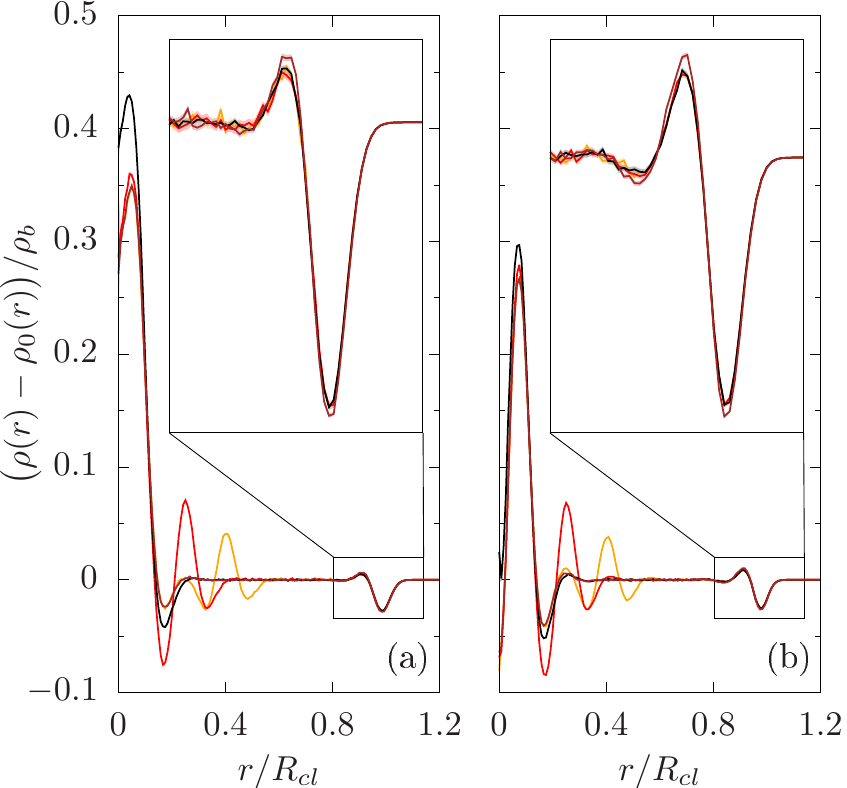}
	\end{adjustbox}
	\vspace{-0.5cm}\caption{\label{fig:densities}Density variation $\rho(r)-\rho_0(r)$ of the QE state with respect to the Laughlin background, normalized to the bulk density $\rho_b=\nu/2\pi$ for (a) $\nu=1/2$ and (b) $\nu=1/3$. In both cases we chose $N=150$.
	Different curves are the QE $\Psi^{[M]}$ [Eq.~\eqref{eq:truncatedQEwavefunction}] densities at (black) $M=0$ (red) $M=9$ (orange) $M=24$ and (brown) $M=149$.
	The shadows of the curves are Monte-Carlo statistical uncertainties.
	Distances have been normalized in units of the cloud's radius.
	The insets are a magnification of the density variation at the edge.}
\end{figure}

\textit{\textbf{Truncated wavefunctions.}}
In this section we introduce a class of wavefunctions that interpolate between JQE and LQE.
In order to do so, we modify Eq.~\eqref{eq:LaughlinCF} by introducing a ``$M$-th orbital truncation": 
\begin{equation}
	\label{eq:truncated_LaughlinCF}
	\Phi^{[M]} (\{z_i \}) =
	\left|
	{
		\begin{array}{cccc}
			\psi_{-1}(z_0) & \psi_{-1}(z_1) & \psi_{-1}(z_2) & \hdots\\
			\vdots & \vdots & \vdots & \hdots \\
			\psi_{M-1}(z_0) & \psi_{M-1}(z_1) & \psi_{M-1}(z_2) & \hdots \\
			z_0^M & z_1^M & z_2^M & \hdots \\
			z_0^{M+1} & z_1^{M+1} & z_2^{M+1} & \hdots \\
			\vdots & \vdots & \vdots & \ddots \\
		\end{array}
	}
	\right|.
\end{equation}
The orbitals in the Slater determinant correspond to those of LQE in Eq.~\eqref{eq:normalizedState} up to the $M$-th. 
From the $(M+1)$-th on, the CF orbitals are those of the Jain's state.
By construction, for $M=N-1$ we recover the LQE [Eq.~\eqref{eq:LaughlinCF}], whereas truncating at $M=0$ gives exactly the JQE [Eq.~\eqref{eq:JainQE}].
A LLL QE wavefunction which interpolates between JQE and LQE can therefore be constructed by ``attaching vortices" to the non-interacting fermions~\cite{jainCompositeFermionsBook_2007} as in standard CF wavefunctions:
\begin{equation}
	\label{eq:truncatedQEwavefunction}
	\psi^{[M]} (\{ z_i \}) = \hat{P}_{LLL}\,\,\Phi^{[M]}(\{z_i\})\,\,\prod_{i<j}(z_i-z_j)^{m-1}.
\end{equation}

Notice that at $M=N-1$, projecting according to the ``standard" Girvin-Jach prescription~\cite{tongLectureNotes, GirvinJach_PRB_1984} leads to Laughlin's version of the QE [Eq.~\eqref{eq:LaughlinQE}].
For a generic truncation orbital $M\gtrsim1$, however, such a projection scheme is intractable.
Since one heuristically expects that the topological features of the state are independent of the projection scheme,
we resort on the Jain-Kamilla method~\cite{JainKamilla_IntJModPhysB_1997,Jain_PRB_1997,jainCompositeFermionsBook_2007}, which in this case is easier to use. In the bosonic case, we employ a modified version of the same scheme~\cite{Chang_PRA_2005} (see Supplementary Materials for technical details~\cite{SM}).
It must be noted that the QE obtained for $M=N-1$ with the Jain-Kamilla scheme does not mathematically coincide with the LQE.
Yet, as it has been argued in Ref.~\cite{Hansson_RMP_2017}, the projection scheme only affects the short-distance physics of the state and thus we do not expect that relying on the Jain-Kamilla will invalidate our analysis.

\textit{\textbf{Numerical analysis.}}
In this section we numerically study the truncated wavefunctions of Eq.~\eqref{eq:truncatedQEwavefunction} by performing standard Monte-Carlo sampling~\cite{Metropolis_1949,Hastings_1970}. 
The wavefunctions have circular symmetry and thus we will present data for their density profile as a function of the radial distance.

First of all, we focus on the variation of the system's density with respect to that of the Laughlin GS, $\rho(r) -\rho_0(r)$. 
We plot such a quantity in Fig.~\ref{fig:densities} for various values of the truncation parameter $M$.
As $M$ initially increases, so does the size of the QE, characterized by large density oscillations;  however, as long as $M\ll N$, the edge is independent of $M$ (see the insets). 
On the other hand, when the CF occupation tail reaches the boundary (namely, for $M=N-1$), the QE shrinks, and indeed the $M=0$ and $M=N-1$ density profiles are qualitatively similar. 
Crucially however, as it can be seen from the insets, the compressible edge is not the same as in the $M\ll N$ cases. 
We attribute this tiny difference to the fact that for $M=N-1$, the orbital of the first Lambda level with angular momentum $N-1$ is occupied, whereas in the other cases it is unoccupied.

\begin{figure}[t]
	\begin{adjustbox}{width=0.5\textwidth, totalheight=\textheight-2\baselineskip,keepaspectratio,center}
		\includegraphics[]{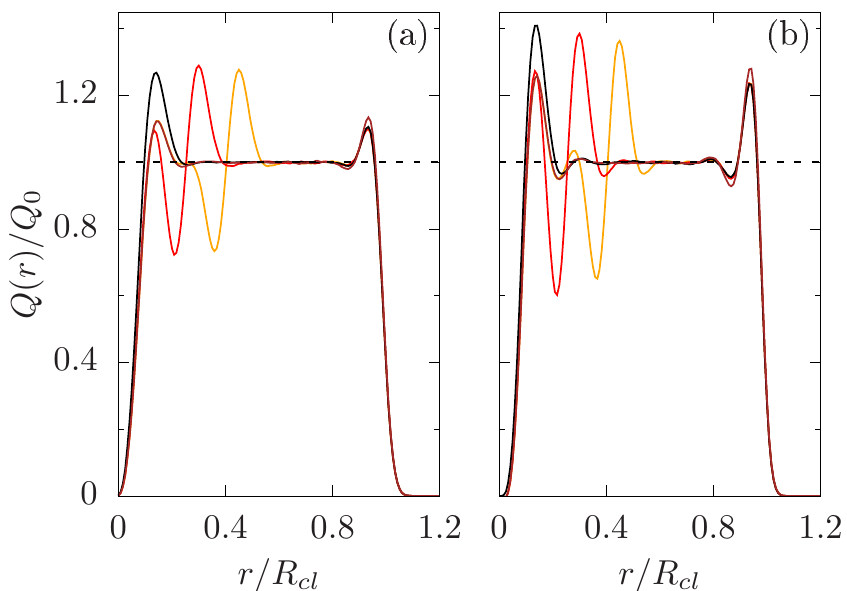}
	\end{adjustbox}
	\vspace{-0.5cm}\caption{\label{fig:charges}QEs charges [Eq.~\eqref{eq:chargeDef}], normalized to the expected one $Q_0=\nu$ for (a) $\nu=1/2$ and (b) $\nu=1/3$. In both cases we chose $N=150$.
		The curves' colours are the same as those of Fig.~\ref{fig:densities}.}
\end{figure}

A second quantity that we study is the  integrated QE excess density with respect to the Laughlin GS:
\begin{equation}
	\label{eq:chargeDef}
	Q(r)=\int_0^r (\rho(r')-\rho_0(r'))d^2r'.
\end{equation}
This is expected to approach the QE charge as $l_B\ll r\ll R_{cl}$, with $R_{cl}=\sqrt{2N/\nu}\,l_B$ being the classical radius of the cloud.
We show our results in Fig.~\ref{fig:charges}. 
It can be seen that the QE charge is independent of the truncation parameter $M$.

\begin{figure}[t]
	\begin{adjustbox}{width=0.5\textwidth, totalheight=\textheight-2\baselineskip,keepaspectratio,center}
		\includegraphics[]{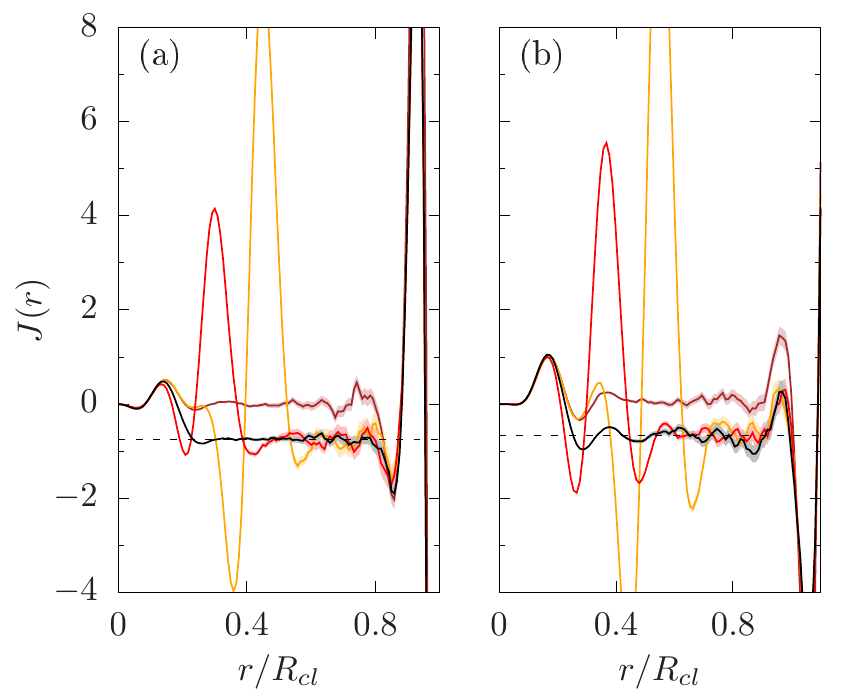}
	\end{adjustbox}
	\vspace{-0.5cm}\caption{\label{fig:spins}QEs spins [Eq.~\eqref{eq:spinDef}], for (a) $\nu=1/2$ and (b) $\nu=1/3$. In both cases we chose $N=150$.
		Black dashed lines denote the expected values in the two cases: $J_{qe}=-3/4$ ($-2/3$) for $\nu=1/2$ ($1/3$).
		The curves' colours are the same as those of Fig.~\ref{fig:densities}.}
\end{figure}

The quantity which sharply distinguishes the $M\sim N$ case from the $M\ll N$ one is the local angular momentum
\begin{equation}
	\label{eq:spinDef}
	J(r)=\int_0^r \left(\frac{{r'}^2}{2l_B^2}-1\right)(\rho(r')-\rho_0(r'))d^2r'
\end{equation}
which for $l_B\ll r\ll R_{cl}$ can be meaningfully associated to a fractional intrinsic anyon spin~\cite{Comparin_2022,Nardin_PRB_2023} and is related to the anyonic braiding properties of the FQHE quasiparticles~\cite{Umucalilar_PhysLettA_2013,Umucalilar_PRL_2018,Umucalilar_PRA_2018,Macaluso_PRL_2019,Macaluso_PRR_2020}.
In particular, the correct value of the spin of a QE that is an anti-anyon of Laughlin's quasihole is~\cite{Einarsson_1995,Gromov_2016,Comparin_2022, Nardin_PRB_2023}
\begin{equation}
	\label{eq:QP_Spin}
	J_{p} = -\frac{p^2}{2m} + \frac{p}{2}.
\end{equation}
Here, $p$ is the charge of the quasiparticle in units of $1/m$. We use the convention $p>0$ for QHs and $p<0$ for QEs (see SM for more details~\cite{SM}).

We plot our numerical results in Fig.~\ref{fig:spins}. 
Let us focus on the value of the spin, corresponding to the plateau appearing for $l_B\ll r\ll R_{cl}$ for all $M$.
The correct value in Eq.~\eqref{eq:QP_Spin} is recovered  only for $M \ll N$, even if the density profiles are actually rather different. In this limit the wavefunctions $\psi^{[M]}$ are good candidate antianyon wavefunctions of Laughlin's QH. 
In the case of the LQE ($M=N-1$), on the other hand, we find a value of the spin which differs from Eq.~\eqref{eq:QP_Spin}.

This behaviour can be traced back to the fact that for $M=N-1$ the density profile of the boundary differs from that of $M \ll N$, as highlighted in Fig.~\ref{fig:densities}. Indeed, all the wavefunctions $\psi^{[M]}$ have the same angular momentum $L=L_0-N$, where $L_0$ is the angular momentum of the Laughlin state. 
By linearity of the angular momentum of the gas {with respect to} the density  $\rho(r)$, it can be shown that $J_\infty =\lim_{r \to \infty} J(r ) = -N$ for all $M$. 
Fig.~\ref{fig:densities} shows clearly that $\rho(r)-\rho_0(r)$ can be split into two parts with disjoint supports: 
one centered around $r=0$ 
and one roughly localised at a distance $r = R_{\rm cl}$. 
The integral $J_\infty$ is thus the sum of two terms: one that corresponds to the spin of the quasiparticle $J_{qp}$, and one that is an edge contribution $J_{\text{edge}}$.
All the wavefunctions which share the same edge profile, as it happens for $M \ll N$, must have the same edge integral $J_{\text{edge}}$, but since $J_{qp}+J_{\text{edge}}=-N$ they also share the same quasiparticle spin.
{Due to the aforementioned non-locality,} for $M=N-1$ the edge density profile is different and consequently $J_{qp}$ and $J_{\text{edge}}$ can take novel values.

At this stage, one could wonder whether this result depends on the choice of using the Jain-Kamilla projection scheme. In Ref.~\cite{Nardin_PRB_2023} we presented a calculation of the LQE spin using the Girvin-Jach projection yielding: $J_{\rm LQE - GJ}=-1/4$ and $J_{\rm LQE-GJ}=0$ at $\nu=1/2$ and $\nu=1/3$, respectively. 
They are not compatible within the statistical uncertainties with those we find in Fig.~\ref{fig:spins}, and they are also incompatible with Eq.~\eqref{eq:QP_Spin}.
We conclude that the LQE spin depends  on the method employed to project out higher Landau levels and does not display any robustness, as instead it should be~\cite{jainCompositeFermionsBook_2007, Hansson_RMP_2017}.

\begin{figure}[t]
	\vspace{0.5cm}
	\begin{adjustbox}{width=0.5\textwidth, totalheight=\textheight-2\baselineskip,keepaspectratio,center}
		\includegraphics[]{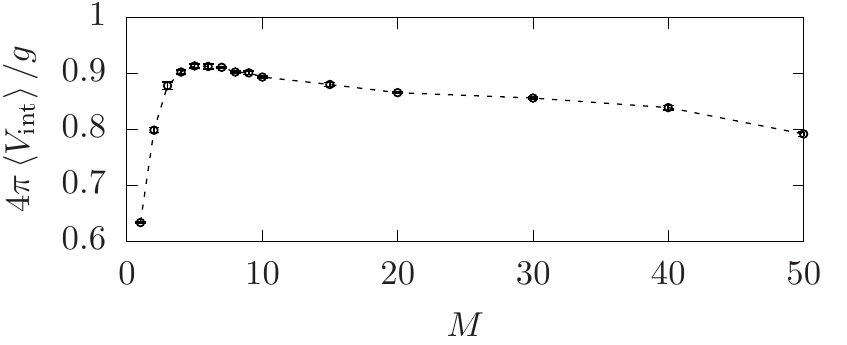}
	\end{adjustbox}
	\vspace{-0.5cm}\caption{Expectation value on the truncated state $\psi^{[M]}$ of $N=50$ bosons at filling $\nu=1/2$ of the contact interaction energy $V_\text{int}=g\sum_{i<j}\delta^{(2)}(\mathbf{r}_i-\mathbf{r}_j)$.
		\label{fig:variationalEnergies}}
\end{figure}

To conclude, let us notice that due to the structure of the truncated LQE in Eq.~\eqref{eq:truncatedQEwavefunction} 
we qualitatively expect (see Fig.~\ref{fig:QE_sketch}) that
as $M$ increases so will the interaction energy of the state: indeed, many more particles come to occupy the first CF Lambda level.
We moreover expect the interaction energy to increase faster at small values of $M$: since $P_1(l)$ decreases at large values of $l$ [Eq.~\eqref{eq:occupationProbabilities}] the first CF Lambda level total occupation will show larger variations when $M$ is small.
This feature qualitatively explains the large variational (Coulomb) energy the LQE has when compared to Jain's proposal, numerically observed in Ref~\citenum{Jeon_2003b}.
We quantitatively validate the aforementioned expectation in Fig.~\ref{fig:variationalEnergies}, where we show the interaction energy expectation value on the truncated state $\psi^{[M]}$ as a function of $M$ in the case of contact interacting bosons at $\nu=1/2$.
It can be seen indeed that the variational interaction energy increases as long as the tail of the QE wavefunction does not touch the system's edge; after this point the energy counter-intuitively slightly decreases again due to the QE contraction -- which can be seen Fig.~\ref{fig:densities} -- being more important than the increased excited Lambda level occupation.
Based on these considerations, we identify these states ($M>0$) as excited states of JQE ($M=0$).

\textit{\textbf{LQEs braiding.}}
As a very final remark, we want to qualitatively comment on the $\mathcal{O}(1/N)$ finite size corrections to the braiding phase of two Laughlin's QE observed in Ref~\citenum{Jeon_2010}. 
We argue that the non-locality we observe can in principle explain such a correction. 
We indeed have shown that when projected onto the LLL the long CF tail affects the compressible edge. 
When two QEs are present we expect not only that they will alter the system's edge in a non-trivial way, but also that they mutually affect each other in a non-local way analogously with what we described for the boundary.
More detailed and quantitative investigations, which go beyond the scope of this letter, are however needed to further address this point.

\textit{\textbf{Conclusions.}}
To summarize, we showed that LQE can be reformulated as a non-local CF with a logarithmic occupation of the first excited Lambda level. 
We defined a class of CF wavefunctions which interpolate between the QEs introduced by Laughlin and by Jain.
We showed that as long as the occupation of the first Lambda level does not affect the system's boundary, the QE spin is robust, and the resulting anyon can be identified as an excited state of Jain's QE which has different interaction energies but the same charge and spin.
On the other hand, we showed that if this tail reaches the system's boundary, the topological protection of the spin is lost, and different LLL projection schemes can lead to different results, meaning that the QE braiding depends on the microscopic structure of the wavefunction. 
This study is based on the notion of fractional spin of FQHE quasiparticles~\cite{Comparin_2022, Nardin_PRB_2023} and constitutes an example of its usefulness in the study of trial wavefunction involving LLL projections.

\textit{\textbf{Acknowledgments}}
We warmly acknowledge discussions with E.~Ardonne on previous related work and with I.~Carusotto.
A.~N.~is supported by \textit{Fondazione Angelo dalla Riccia} and thanks LPTMS for warm hospitality.
This work is supported by LabEx PALM (ANR-10-LABX-0039-PALM) in Orsay, by Region Ile-de-France in the framework of
the DIM Sirteq.


\begin{thebibliography}{47}%
	\makeatletter
	\providecommand \@ifxundefined [1]{%
		\@ifx{#1\undefined}
	}%
	\providecommand \@ifnum [1]{%
		\ifnum #1\expandafter \@firstoftwo
		\else \expandafter \@secondoftwo
		\fi
	}%
	\providecommand \@ifx [1]{%
		\ifx #1\expandafter \@firstoftwo
		\else \expandafter \@secondoftwo
		\fi
	}%
	\providecommand \natexlab [1]{#1}%
	\providecommand \enquote  [1]{``#1''}%
	\providecommand \bibnamefont  [1]{#1}%
	\providecommand \bibfnamefont [1]{#1}%
	\providecommand \citenamefont [1]{#1}%
	\providecommand \href@noop [0]{\@secondoftwo}%
	\providecommand \href [0]{\begingroup \@sanitize@url \@href}%
	\providecommand \@href[1]{\@@startlink{#1}\@@href}%
	\providecommand \@@href[1]{\endgroup#1\@@endlink}%
	\providecommand \@sanitize@url [0]{\catcode `\\12\catcode `\$12\catcode
		`\&12\catcode `\#12\catcode `\^12\catcode `\_12\catcode `\%12\relax}%
	\providecommand \@@startlink[1]{}%
	\providecommand \@@endlink[0]{}%
	\providecommand \url  [0]{\begingroup\@sanitize@url \@url }%
	\providecommand \@url [1]{\endgroup\@href {#1}{\urlprefix }}%
	\providecommand \urlprefix  [0]{URL }%
	\providecommand \Eprint [0]{\href }%
	\providecommand \doibase [0]{https://doi.org/}%
	\providecommand \selectlanguage [0]{\@gobble}%
	\providecommand \bibinfo  [0]{\@secondoftwo}%
	\providecommand \bibfield  [0]{\@secondoftwo}%
	\providecommand \translation [1]{[#1]}%
	\providecommand \BibitemOpen [0]{}%
	\providecommand \bibitemStop [0]{}%
	\providecommand \bibitemNoStop [0]{.\EOS\space}%
	\providecommand \EOS [0]{\spacefactor3000\relax}%
	\providecommand \BibitemShut  [1]{\csname bibitem#1\endcsname}%
	\let\auto@bib@innerbib\@empty
	\bibitem [{\citenamefont {Laughlin}(1983)}]{Laughlin_1983}%
	\BibitemOpen
	\bibfield  {author} {\bibinfo {author} {\bibfnamefont {R.~B.}\ \bibnamefont
			{Laughlin}},\ }\bibfield  {title} {\bibinfo {title} {Anomalous quantum hall
			effect: An incompressible quantum fluid with fractionally charged
			excitations},\ }\href {https://doi.org/10.1103/PhysRevLett.50.1395}
	{\bibfield  {journal} {\bibinfo  {journal} {Phys. Rev. Lett.}\ }\textbf
		{\bibinfo {volume} {50}},\ \bibinfo {pages} {1395} (\bibinfo {year}
		{1983})}\BibitemShut {NoStop}%
	\bibitem [{\citenamefont {Wilczek}\ and\ \citenamefont
		{Shapere}(1989)}]{WilczekShapere_GeometricPhases_1989}%
	\BibitemOpen
	\bibfield  {author} {\bibinfo {author} {\bibfnamefont {F.}~\bibnamefont
			{Wilczek}}\ and\ \bibinfo {author} {\bibfnamefont {A.}~\bibnamefont
			{Shapere}},\ }\href {https://doi.org/10.1142/0613} {\emph {\bibinfo {title}
			{Geometric Phases in Physics}}}\ (\bibinfo  {publisher} {WORLD SCIENTIFIC},\
	\bibinfo {year} {1989})\ \Eprint
	{https://arxiv.org/abs/https://www.worldscientific.com/doi/pdf/10.1142/0613}
	{https://www.worldscientific.com/doi/pdf/10.1142/0613} \BibitemShut {NoStop}%
	\bibitem [{\citenamefont {Arovas}\ \emph {et~al.}(1984)\citenamefont {Arovas},
		\citenamefont {Schrieffer},\ and\ \citenamefont {Wilczek}}]{Arovas_1984}%
	\BibitemOpen
	\bibfield  {author} {\bibinfo {author} {\bibfnamefont {D.}~\bibnamefont
			{Arovas}}, \bibinfo {author} {\bibfnamefont {J.~R.}\ \bibnamefont
			{Schrieffer}},\ and\ \bibinfo {author} {\bibfnamefont {F.}~\bibnamefont
			{Wilczek}},\ }\bibfield  {title} {\bibinfo {title} {Fractional statistics and
			the quantum hall effect},\ }\href
	{https://doi.org/10.1103/PhysRevLett.53.722} {\bibfield  {journal} {\bibinfo
			{journal} {Phys. Rev. Lett.}\ }\textbf {\bibinfo {volume} {53}},\ \bibinfo
		{pages} {722} (\bibinfo {year} {1984})}\BibitemShut {NoStop}%
	\bibitem [{\citenamefont {{Kj{\o}nsberg}}\ and\ \citenamefont
		{{Leinaas}}(1997)}]{KjonsberLeinaas_IntJModPhysA_1997}%
	\BibitemOpen
	\bibfield  {author} {\bibinfo {author} {\bibfnamefont {H.}~\bibnamefont
			{{Kj{\o}nsberg}}}\ and\ \bibinfo {author} {\bibfnamefont {J.~M.}\
			\bibnamefont {{Leinaas}}},\ }\bibfield  {title} {\bibinfo {title} {{On the
				Anyon Description of the Laughlin Hole States}},\ }\href
	{https://doi.org/10.1142/S0217751X97001250} {\bibfield  {journal} {\bibinfo
			{journal} {International Journal of Modern Physics A}\ }\textbf {\bibinfo
			{volume} {12}},\ \bibinfo {pages} {1975} (\bibinfo {year} {1997})},\ \Eprint
	{https://arxiv.org/abs/cond-mat/9606214} {arXiv:cond-mat/9606214 [cond-mat]}
	\BibitemShut {NoStop}%
	\bibitem [{\citenamefont {Kjønsberg}\ and\ \citenamefont
		{Leinaas}(1999)}]{KJONSBERG1999705}%
	\BibitemOpen
	\bibfield  {author} {\bibinfo {author} {\bibfnamefont {H.}~\bibnamefont
			{Kjønsberg}}\ and\ \bibinfo {author} {\bibfnamefont {J.}~\bibnamefont
			{Leinaas}},\ }\bibfield  {title} {\bibinfo {title} {Charge and statistics of
			quantum hall quasi-particles — a numerical study of mean values and
			fluctuations},\ }\href
	{https://doi.org/https://doi.org/10.1016/S0550-3213(99)00353-3} {\bibfield
		{journal} {\bibinfo  {journal} {Nuclear Physics B}\ }\textbf {\bibinfo
			{volume} {559}},\ \bibinfo {pages} {705} (\bibinfo {year}
		{1999})}\BibitemShut {NoStop}%
	\bibitem [{\citenamefont {Kj{\o}nsberg}\ and\ \citenamefont
		{Myrheim}(1999)}]{KJONSBERG_1999}%
	\BibitemOpen
	\bibfield  {author} {\bibinfo {author} {\bibfnamefont {H.}~\bibnamefont
			{Kj{\o}nsberg}}\ and\ \bibinfo {author} {\bibfnamefont {J.}~\bibnamefont
			{Myrheim}},\ }\bibfield  {title} {\bibinfo {title} {Numerical study of charge
			and statistics of laughlin quasiparticles},\ }\href
	{https://doi.org/10.1142/s0217751x99000270} {\bibfield  {journal} {\bibinfo
			{journal} {International Journal of Modern Physics A}\ }\textbf {\bibinfo
			{volume} {14}},\ \bibinfo {pages} {537} (\bibinfo {year} {1999})}\BibitemShut
	{NoStop}%
	\bibitem [{\citenamefont {Jeon}\ and\ \citenamefont {Jain}(2010)}]{Jeon_2010}%
	\BibitemOpen
	\bibfield  {author} {\bibinfo {author} {\bibfnamefont {G.~S.}\ \bibnamefont
			{Jeon}}\ and\ \bibinfo {author} {\bibfnamefont {J.~K.}\ \bibnamefont
			{Jain}},\ }\bibfield  {title} {\bibinfo {title} {Thermodynamic behavior of
			braiding statistics for certain fractional quantum hall quasiparticles},\
	}\href {https://doi.org/10.1103/PhysRevB.81.035319} {\bibfield  {journal}
		{\bibinfo  {journal} {Phys. Rev. B}\ }\textbf {\bibinfo {volume} {81}},\
		\bibinfo {pages} {035319} (\bibinfo {year} {2010})}\BibitemShut {NoStop}%
	\bibitem [{\citenamefont {Wilczek}(1982{\natexlab{a}})}]{Wilczek_PRL_1982a}%
	\BibitemOpen
	\bibfield  {author} {\bibinfo {author} {\bibfnamefont {F.}~\bibnamefont
			{Wilczek}},\ }\bibfield  {title} {\bibinfo {title} {Magnetic flux, angular
			momentum, and statistics},\ }\href
	{https://doi.org/10.1103/PhysRevLett.48.1144} {\bibfield  {journal} {\bibinfo
			{journal} {Phys. Rev. Lett.}\ }\textbf {\bibinfo {volume} {48}},\ \bibinfo
		{pages} {1144} (\bibinfo {year} {1982}{\natexlab{a}})}\BibitemShut {NoStop}%
	\bibitem [{\citenamefont {Wilczek}(1982{\natexlab{b}})}]{Wilczek_PRL_1982b}%
	\BibitemOpen
	\bibfield  {author} {\bibinfo {author} {\bibfnamefont {F.}~\bibnamefont
			{Wilczek}},\ }\bibfield  {title} {\bibinfo {title} {Magnetic flux, angular
			momentum, and statistics},\ }\href
	{https://doi.org/10.1103/PhysRevLett.48.1144} {\bibfield  {journal} {\bibinfo
			{journal} {Phys. Rev. Lett.}\ }\textbf {\bibinfo {volume} {48}},\ \bibinfo
		{pages} {1144} (\bibinfo {year} {1982}{\natexlab{b}})}\BibitemShut {NoStop}%
	\bibitem [{\citenamefont {Preskill}(2004)}]{Preskill_2004}%
	\BibitemOpen
	\bibfield  {author} {\bibinfo {author} {\bibfnamefont {J.}~\bibnamefont
			{Preskill}},\ }\href
	{http://www.theory.caltech.edu/~preskill/ph219/topological.pdf} {\emph
		{\bibinfo {title} {Lecture notes Lecture Notes for Physics 219: Quantum
				Computation}}}\ (\bibinfo {year} {2004})\ Chap.\ \bibinfo {chapter} {9:
		Topological quantum computation}\BibitemShut {NoStop}%
	\bibitem [{\citenamefont {Li}(1992)}]{Li_1992}%
	\BibitemOpen
	\bibfield  {author} {\bibinfo {author} {\bibfnamefont {D.}~\bibnamefont
			{Li}},\ }\bibfield  {title} {\bibinfo {title} {The spin of the quasi-particle
			in the fractional quantum hall effect},\ }\href
	{https://doi.org/https://doi.org/10.1016/0375-9601(92)90810-9} {\bibfield
		{journal} {\bibinfo  {journal} {Phys. Lett. A}\ }\textbf {\bibinfo {volume}
			{169}},\ \bibinfo {pages} {82} (\bibinfo {year} {1992})}\BibitemShut
	{NoStop}%
	\bibitem [{\citenamefont {Li}(1993)}]{Li_1993}%
	\BibitemOpen
	\bibfield  {author} {\bibinfo {author} {\bibfnamefont {D.}~\bibnamefont
			{Li}},\ }\bibfield  {title} {\bibinfo {title} {Intrinsic quasiparticle's spin
			and fractional quantum hall effect on riemann surfaces},\ }\href
	{https://doi.org/10.1142/S0217984993001090} {\bibfield  {journal} {\bibinfo
			{journal} {Mod. Phys. Lett. B}\ }\textbf {\bibinfo {volume} {07}},\ \bibinfo
		{pages} {1103} (\bibinfo {year} {1993})}\BibitemShut {NoStop}%
	\bibitem [{\citenamefont {Einarsson}\ \emph {et~al.}(1995)\citenamefont
		{Einarsson}, \citenamefont {Sondhi}, \citenamefont {Girvin},\ and\
		\citenamefont {Arovas}}]{Einarsson_1995}%
	\BibitemOpen
	\bibfield  {author} {\bibinfo {author} {\bibfnamefont {T.}~\bibnamefont
			{Einarsson}}, \bibinfo {author} {\bibfnamefont {S.~L.}\ \bibnamefont
			{Sondhi}}, \bibinfo {author} {\bibfnamefont {S.~M.}\ \bibnamefont {Girvin}},\
		and\ \bibinfo {author} {\bibfnamefont {D.~P.}\ \bibnamefont {Arovas}},\
	}\bibfield  {title} {\bibinfo {title} {Fractional spin for quantum hall
			effect quasiparticles},\ }\href
	{https://doi.org/https://doi.org/10.1016/0550-3213(95)00025-N} {\bibfield
		{journal} {\bibinfo  {journal} {Nucl. Phys. B}\ }\textbf {\bibinfo {volume}
			{441}},\ \bibinfo {pages} {515} (\bibinfo {year} {1995})}\BibitemShut
	{NoStop}%
	\bibitem [{\citenamefont {Read}(2008)}]{Read_Arxiv2008_Spin}%
	\BibitemOpen
	\bibfield  {author} {\bibinfo {author} {\bibfnamefont {N.}~\bibnamefont
			{Read}},\ }\href@noop {} {\bibinfo {title} {Quasiparticle spin from adiabatic
			transport in quantum hall trial wavefunctions}} (\bibinfo {year} {2008}),\
	\Eprint {https://arxiv.org/abs/0807.3107} {arXiv:0807.3107
		[cond-mat.mes-hall]} \BibitemShut {NoStop}%
	\bibitem [{\citenamefont {Gromov}(2016)}]{Gromov_2016}%
	\BibitemOpen
	\bibfield  {author} {\bibinfo {author} {\bibfnamefont {A.}~\bibnamefont
			{Gromov}},\ }\bibfield  {title} {\bibinfo {title} {Geometric defects in
			quantum hall states},\ }\href {https://doi.org/10.1103/PhysRevB.94.085116}
	{\bibfield  {journal} {\bibinfo  {journal} {Phys. Rev. B}\ }\textbf {\bibinfo
			{volume} {94}},\ \bibinfo {pages} {085116} (\bibinfo {year}
		{2016})}\BibitemShut {NoStop}%
	\bibitem [{\citenamefont {Trung}\ \emph {et~al.}(2023)\citenamefont {Trung},
		\citenamefont {Wang},\ and\ \citenamefont {Yang}}]{Trung_PRB_2023}%
	\BibitemOpen
	\bibfield  {author} {\bibinfo {author} {\bibfnamefont {H.~Q.}\ \bibnamefont
			{Trung}}, \bibinfo {author} {\bibfnamefont {Y.}~\bibnamefont {Wang}},\ and\
		\bibinfo {author} {\bibfnamefont {B.}~\bibnamefont {Yang}},\ }\bibfield
	{title} {\bibinfo {title} {Spin-statistics relation and abelian braiding
			phase for anyons in the fractional quantum hall effect},\ }\bibfield
	{journal} {\bibinfo  {journal} {Physical Review B}\ }\textbf {\bibinfo
		{volume} {107}},\ \href {https://doi.org/10.1103/physrevb.107.l201301}
	{10.1103/physrevb.107.l201301} (\bibinfo {year} {2023})\BibitemShut {NoStop}%
	\bibitem [{\citenamefont {Macaluso}\ \emph {et~al.}(2020)\citenamefont
		{Macaluso}, \citenamefont {Comparin}, \citenamefont {Umucal\ifmmode \imath
			\else~\i \fi{}lar}, \citenamefont {Gerster}, \citenamefont {Montangero},
		\citenamefont {Rizzi},\ and\ \citenamefont {Carusotto}}]{Macaluso_PRR_2020}%
	\BibitemOpen
	\bibfield  {author} {\bibinfo {author} {\bibfnamefont {E.}~\bibnamefont
			{Macaluso}}, \bibinfo {author} {\bibfnamefont {T.}~\bibnamefont {Comparin}},
		\bibinfo {author} {\bibfnamefont {R.~O.}\ \bibnamefont {Umucal\ifmmode \imath
				\else~\i \fi{}lar}}, \bibinfo {author} {\bibfnamefont {M.}~\bibnamefont
			{Gerster}}, \bibinfo {author} {\bibfnamefont {S.}~\bibnamefont {Montangero}},
		\bibinfo {author} {\bibfnamefont {M.}~\bibnamefont {Rizzi}},\ and\ \bibinfo
		{author} {\bibfnamefont {I.}~\bibnamefont {Carusotto}},\ }\bibfield  {title}
	{\bibinfo {title} {Charge and statistics of lattice quasiholes from density
			measurements: A tree tensor network study},\ }\href
	{https://doi.org/10.1103/PhysRevResearch.2.013145} {\bibfield  {journal}
		{\bibinfo  {journal} {Phys. Rev. Res.}\ }\textbf {\bibinfo {volume} {2}},\
		\bibinfo {pages} {013145} (\bibinfo {year} {2020})}\BibitemShut {NoStop}%
	\bibitem [{\citenamefont {Comparin}\ \emph {et~al.}(2022)\citenamefont
		{Comparin}, \citenamefont {Opler}, \citenamefont {Macaluso}, \citenamefont
		{Biella}, \citenamefont {Polychronakos},\ and\ \citenamefont
		{Mazza}}]{Comparin_2022}%
	\BibitemOpen
	\bibfield  {author} {\bibinfo {author} {\bibfnamefont {T.}~\bibnamefont
			{Comparin}}, \bibinfo {author} {\bibfnamefont {A.}~\bibnamefont {Opler}},
		\bibinfo {author} {\bibfnamefont {E.}~\bibnamefont {Macaluso}}, \bibinfo
		{author} {\bibfnamefont {A.}~\bibnamefont {Biella}}, \bibinfo {author}
		{\bibfnamefont {A.~P.}\ \bibnamefont {Polychronakos}},\ and\ \bibinfo
		{author} {\bibfnamefont {L.}~\bibnamefont {Mazza}},\ }\bibfield  {title}
	{\bibinfo {title} {Measurable fractional spin for quantum hall quasiparticles
			on the disk},\ }\href {https://doi.org/10.1103/PhysRevB.105.085125}
	{\bibfield  {journal} {\bibinfo  {journal} {Phys. Rev. B}\ }\textbf {\bibinfo
			{volume} {105}},\ \bibinfo {pages} {085125} (\bibinfo {year}
		{2022})}\BibitemShut {NoStop}%
	\bibitem [{\citenamefont {Nardin}\ \emph {et~al.}(2022)\citenamefont {Nardin},
		\citenamefont {Ardonne},\ and\ \citenamefont {Mazza}}]{Nardin_PRB_2023}%
	\BibitemOpen
	\bibfield  {author} {\bibinfo {author} {\bibfnamefont {A.}~\bibnamefont
			{Nardin}}, \bibinfo {author} {\bibfnamefont {E.}~\bibnamefont {Ardonne}},\
		and\ \bibinfo {author} {\bibfnamefont {L.}~\bibnamefont {Mazza}},\
	}\href@noop {} {\bibinfo {title} {Spin-statistics relation for abelian
			quantum hall states}} (\bibinfo {year} {2022}),\ \Eprint
	{https://arxiv.org/abs/2211.07788} {arXiv:2211.07788 [cond-mat.mes-hall]}
	\BibitemShut {NoStop}%
	\bibitem [{\citenamefont {Goerbig}(2009)}]{goerbig2009quantum}%
	\BibitemOpen
	\bibfield  {author} {\bibinfo {author} {\bibfnamefont {M.~O.}\ \bibnamefont
			{Goerbig}},\ }\href@noop {} {\bibinfo {title} {Quantum hall effects}}
	(\bibinfo {year} {2009}),\ \Eprint {https://arxiv.org/abs/0909.1998}
	{arXiv:0909.1998 [cond-mat.mes-hall]} \BibitemShut {NoStop}%
	\bibitem [{\citenamefont {Tong}(2016)}]{tongLectureNotes}%
	\BibitemOpen
	\bibfield  {author} {\bibinfo {author} {\bibfnamefont {D.}~\bibnamefont
			{Tong}},\ }\bibfield  {title} {\bibinfo {title} {Lectures on the quantum hall
			effect}\ }\href {https://doi.org/10.48550/ARXIV.1606.06687}
	{10.48550/ARXIV.1606.06687} (\bibinfo {year} {2016})\BibitemShut {NoStop}%
	\bibitem [{\citenamefont {Simon}(2016)}]{simon2016topological}%
	\BibitemOpen
	\bibfield  {author} {\bibinfo {author} {\bibfnamefont {S.}~\bibnamefont
			{Simon}},\ }\href@noop {} {\bibinfo {title} {Topological quantum: Lecture
			notes}} (\bibinfo {year} {2016})\BibitemShut {NoStop}%
	\bibitem [{\citenamefont {Jain}(1989)}]{Jain_1989}%
	\BibitemOpen
	\bibfield  {author} {\bibinfo {author} {\bibfnamefont {J.~K.}\ \bibnamefont
			{Jain}},\ }\bibfield  {title} {\bibinfo {title} {Composite-fermion approach
			for the fractional quantum hall effect},\ }\href
	{https://doi.org/10.1103/PhysRevLett.63.199} {\bibfield  {journal} {\bibinfo
			{journal} {Phys. Rev. Lett.}\ }\textbf {\bibinfo {volume} {63}},\ \bibinfo
		{pages} {199} (\bibinfo {year} {1989})}\BibitemShut {NoStop}%
	\bibitem [{\citenamefont {Jain}(2007)}]{jainCompositeFermionsBook_2007}%
	\BibitemOpen
	\bibfield  {author} {\bibinfo {author} {\bibfnamefont {J.~K.}\ \bibnamefont
			{Jain}},\ }\href {https://doi.org/10.1017/CBO9780511607561} {\emph {\bibinfo
			{title} {Composite Fermions}}}\ (\bibinfo  {publisher} {Cambridge University
		Press},\ \bibinfo {year} {2007})\BibitemShut {NoStop}%
	\bibitem [{\citenamefont {Jeon}\ \emph {et~al.}(2003)\citenamefont {Jeon},
		\citenamefont {Graham},\ and\ \citenamefont {Jain}}]{Jeon_2003}%
	\BibitemOpen
	\bibfield  {author} {\bibinfo {author} {\bibfnamefont {G.~S.}\ \bibnamefont
			{Jeon}}, \bibinfo {author} {\bibfnamefont {K.~L.}\ \bibnamefont {Graham}},\
		and\ \bibinfo {author} {\bibfnamefont {J.~K.}\ \bibnamefont {Jain}},\
	}\bibfield  {title} {\bibinfo {title} {Fractional statistics in the
			fractional quantum hall effect},\ }\href
	{https://doi.org/10.1103/PhysRevLett.91.036801} {\bibfield  {journal}
		{\bibinfo  {journal} {Phys. Rev. Lett.}\ }\textbf {\bibinfo {volume} {91}},\
		\bibinfo {pages} {036801} (\bibinfo {year} {2003})}\BibitemShut {NoStop}%
	\bibitem [{\citenamefont {Jeon}\ \emph {et~al.}(2004)\citenamefont {Jeon},
		\citenamefont {Graham},\ and\ \citenamefont {Jain}}]{Jeon_2004}%
	\BibitemOpen
	\bibfield  {author} {\bibinfo {author} {\bibfnamefont {G.~S.}\ \bibnamefont
			{Jeon}}, \bibinfo {author} {\bibfnamefont {K.~L.}\ \bibnamefont {Graham}},\
		and\ \bibinfo {author} {\bibfnamefont {J.~K.}\ \bibnamefont {Jain}},\
	}\bibfield  {title} {\bibinfo {title} {Berry phases for composite fermions:
			Effective magnetic field and fractional statistics},\ }\href
	{https://doi.org/10.1103/PhysRevB.70.125316} {\bibfield  {journal} {\bibinfo
			{journal} {Phys. Rev. B}\ }\textbf {\bibinfo {volume} {70}},\ \bibinfo
		{pages} {125316} (\bibinfo {year} {2004})}\BibitemShut {NoStop}%
	\bibitem [{\citenamefont {Hansson}\ \emph {et~al.}(2007)\citenamefont
		{Hansson}, \citenamefont {Chang}, \citenamefont {Jain},\ and\ \citenamefont
		{Viefers}}]{Hansson_PRB_2007}%
	\BibitemOpen
	\bibfield  {author} {\bibinfo {author} {\bibfnamefont {T.~H.}\ \bibnamefont
			{Hansson}}, \bibinfo {author} {\bibfnamefont {C.-C.}\ \bibnamefont {Chang}},
		\bibinfo {author} {\bibfnamefont {J.~K.}\ \bibnamefont {Jain}},\ and\
		\bibinfo {author} {\bibfnamefont {S.}~\bibnamefont {Viefers}},\ }\bibfield
	{title} {\bibinfo {title} {Composite-fermion wave functions as correlators in
			conformal field theory},\ }\href {https://doi.org/10.1103/PhysRevB.76.075347}
	{\bibfield  {journal} {\bibinfo  {journal} {Phys. Rev. B}\ }\textbf {\bibinfo
			{volume} {76}},\ \bibinfo {pages} {075347} (\bibinfo {year}
		{2007})}\BibitemShut {NoStop}%
	\bibitem [{\citenamefont {Kjäll}\ \emph {et~al.}(2018)\citenamefont {Kjäll},
		\citenamefont {Ardonne}, \citenamefont {Dwivedi}, \citenamefont {Hermanns},\
		and\ \citenamefont {Hansson}}]{Kjall_2018}%
	\BibitemOpen
	\bibfield  {author} {\bibinfo {author} {\bibfnamefont {J.}~\bibnamefont
			{Kjäll}}, \bibinfo {author} {\bibfnamefont {E.}~\bibnamefont {Ardonne}},
		\bibinfo {author} {\bibfnamefont {V.}~\bibnamefont {Dwivedi}}, \bibinfo
		{author} {\bibfnamefont {M.}~\bibnamefont {Hermanns}},\ and\ \bibinfo
		{author} {\bibfnamefont {T.~H.}\ \bibnamefont {Hansson}},\ }\bibfield
	{title} {\bibinfo {title} {Matrix product state representation of
			quasielectron wave functions},\ }\href
	{https://doi.org/10.1088/1742-5468/aab679} {\bibfield  {journal} {\bibinfo
			{journal} {J. Stat. Mech.: Theory Exp.}\ }\textbf {\bibinfo {volume}
			{2018}}\bibinfo  {number} { (5)},\ \bibinfo {pages} {053101}}\BibitemShut
	{NoStop}%
	\bibitem [{\citenamefont {Hansson}\ \emph {et~al.}(2009)\citenamefont
		{Hansson}, \citenamefont {Hermanns}, \citenamefont {Regnault},\ and\
		\citenamefont {Viefers}}]{Hansson_PRL_2009}%
	\BibitemOpen
	\bibfield  {number} {  }\bibfield  {author} {\bibinfo {author} {\bibfnamefont
			{T.~H.}\ \bibnamefont {Hansson}}, \bibinfo {author} {\bibfnamefont
			{M.}~\bibnamefont {Hermanns}}, \bibinfo {author} {\bibfnamefont
			{N.}~\bibnamefont {Regnault}},\ and\ \bibinfo {author} {\bibfnamefont
			{S.}~\bibnamefont {Viefers}},\ }\bibfield  {title} {\bibinfo {title}
		{Conformal field theory approach to abelian and non-abelian quantum hall
			quasielectrons},\ }\href {https://doi.org/10.1103/PhysRevLett.102.166805}
	{\bibfield  {journal} {\bibinfo  {journal} {Phys. Rev. Lett.}\ }\textbf
		{\bibinfo {volume} {102}},\ \bibinfo {pages} {166805} (\bibinfo {year}
		{2009})}\BibitemShut {NoStop}%
	\bibitem [{\citenamefont {Dev}\ and\ \citenamefont
		{Jain}(1992)}]{Dev_PRB_1992}%
	\BibitemOpen
	\bibfield  {author} {\bibinfo {author} {\bibfnamefont {G.}~\bibnamefont
			{Dev}}\ and\ \bibinfo {author} {\bibfnamefont {J.~K.}\ \bibnamefont {Jain}},\
	}\bibfield  {title} {\bibinfo {title} {Jastrow-slater trial wave functions
			for the fractional quantum hall effect: Results for few-particle systems},\
	}\href {https://doi.org/10.1103/PhysRevB.45.1223} {\bibfield  {journal}
		{\bibinfo  {journal} {Phys. Rev. B}\ }\textbf {\bibinfo {volume} {45}},\
		\bibinfo {pages} {1223} (\bibinfo {year} {1992})}\BibitemShut {NoStop}%
	\bibitem [{\citenamefont {Jeon}\ and\ \citenamefont {Jain}(2003)}]{Jeon_2003b}%
	\BibitemOpen
	\bibfield  {author} {\bibinfo {author} {\bibfnamefont {G.~S.}\ \bibnamefont
			{Jeon}}\ and\ \bibinfo {author} {\bibfnamefont {J.~K.}\ \bibnamefont
			{Jain}},\ }\bibfield  {title} {\bibinfo {title} {Nature of quasiparticle
			excitations in the fractional quantum hall effect},\ }\href
	{https://doi.org/10.1103/PhysRevB.68.165346} {\bibfield  {journal} {\bibinfo
			{journal} {Phys. Rev. B}\ }\textbf {\bibinfo {volume} {68}},\ \bibinfo
		{pages} {165346} (\bibinfo {year} {2003})}\BibitemShut {NoStop}%
	\bibitem [{\citenamefont {Bernevig}\ and\ \citenamefont
		{Haldane}(2009)}]{Bernevig_PRL_2009}%
	\BibitemOpen
	\bibfield  {author} {\bibinfo {author} {\bibfnamefont {B.~A.}\ \bibnamefont
			{Bernevig}}\ and\ \bibinfo {author} {\bibfnamefont {F.~D.~M.}\ \bibnamefont
			{Haldane}},\ }\bibfield  {title} {\bibinfo {title} {Clustering properties and
			model wave functions for non-abelian fractional quantum hall
			quasielectrons},\ }\href {https://doi.org/10.1103/PhysRevLett.102.066802}
	{\bibfield  {journal} {\bibinfo  {journal} {Phys. Rev. Lett.}\ }\textbf
		{\bibinfo {volume} {102}},\ \bibinfo {pages} {066802} (\bibinfo {year}
		{2009})}\BibitemShut {NoStop}%
	\bibitem [{\citenamefont {{Bochniak}}\ \emph {et~al.}(2022)\citenamefont
		{{Bochniak}}, \citenamefont {{Nussinov}}, \citenamefont {{Seidel}},\ and\
		\citenamefont {{Ortiz}}}]{Bochniak_CommPhys_2022}%
	\BibitemOpen
	\bibfield  {author} {\bibinfo {author} {\bibfnamefont {A.}~\bibnamefont
			{{Bochniak}}}, \bibinfo {author} {\bibfnamefont {Z.}~\bibnamefont
			{{Nussinov}}}, \bibinfo {author} {\bibfnamefont {A.}~\bibnamefont
			{{Seidel}}},\ and\ \bibinfo {author} {\bibfnamefont {G.}~\bibnamefont
			{{Ortiz}}},\ }\bibfield  {title} {\bibinfo {title} {{Mechanism for particle
				fractionalization and universal edge physics in quantum Hall fluids}},\
	}\href {https://doi.org/10.1038/s42005-022-00946-8} {\bibfield  {journal}
		{\bibinfo  {journal} {Communications Physics}\ }\textbf {\bibinfo {volume}
			{5}},\ \bibinfo {eid} {171} (\bibinfo {year} {2022})},\ \Eprint
	{https://arxiv.org/abs/2110.06216} {arXiv:2110.06216 [cond-mat.str-el]}
	\BibitemShut {NoStop}%
	\bibitem [{\citenamefont {Bochniak}\ and\ \citenamefont
		{Ortiz}(2023)}]{bochniak2023fusion}%
	\BibitemOpen
	\bibfield  {author} {\bibinfo {author} {\bibfnamefont {A.}~\bibnamefont
			{Bochniak}}\ and\ \bibinfo {author} {\bibfnamefont {G.}~\bibnamefont
			{Ortiz}},\ }\href@noop {} {\bibinfo {title} {Fusion mechanism for
			quasiparticles and topological quantum order in the lowest-landau-level}}
	(\bibinfo {year} {2023}),\ \Eprint {https://arxiv.org/abs/2308.03548}
	{arXiv:2308.03548 [cond-mat.str-el]} \BibitemShut {NoStop}%
	\bibitem [{\citenamefont {Girvin}\ and\ \citenamefont
		{Jach}(1984)}]{GirvinJach_PRB_1984}%
	\BibitemOpen
	\bibfield  {author} {\bibinfo {author} {\bibfnamefont {S.~M.}\ \bibnamefont
			{Girvin}}\ and\ \bibinfo {author} {\bibfnamefont {T.}~\bibnamefont {Jach}},\
	}\bibfield  {title} {\bibinfo {title} {Formalism for the quantum hall effect:
			Hilbert space of analytic functions},\ }\href
	{https://doi.org/10.1103/PhysRevB.29.5617} {\bibfield  {journal} {\bibinfo
			{journal} {Phys. Rev. B}\ }\textbf {\bibinfo {volume} {29}},\ \bibinfo
		{pages} {5617} (\bibinfo {year} {1984})}\BibitemShut {NoStop}%
	\bibitem [{\citenamefont {Jain}\ and\ \citenamefont
		{Kamilla}(1997{\natexlab{a}})}]{JainKamilla_IntJModPhysB_1997}%
	\BibitemOpen
	\bibfield  {author} {\bibinfo {author} {\bibfnamefont {J.~K.}\ \bibnamefont
			{Jain}}\ and\ \bibinfo {author} {\bibfnamefont {R.~K.}\ \bibnamefont
			{Kamilla}},\ }\bibfield  {title} {\bibinfo {title} {Composite fermions in the
			hilbert space of the lowest electronic landau level},\ }\href
	{https://doi.org/10.1142/S0217979297001301} {\bibfield  {journal} {\bibinfo
			{journal} {International Journal of Modern Physics B}\ }\textbf {\bibinfo
			{volume} {11}},\ \bibinfo {pages} {2621} (\bibinfo {year}
		{1997}{\natexlab{a}})},\ \Eprint
	{https://arxiv.org/abs/https://doi.org/10.1142/S0217979297001301}
	{https://doi.org/10.1142/S0217979297001301} \BibitemShut {NoStop}%
	\bibitem [{\citenamefont {Jain}\ and\ \citenamefont
		{Kamilla}(1997{\natexlab{b}})}]{Jain_PRB_1997}%
	\BibitemOpen
	\bibfield  {author} {\bibinfo {author} {\bibfnamefont {J.~K.}\ \bibnamefont
			{Jain}}\ and\ \bibinfo {author} {\bibfnamefont {R.~K.}\ \bibnamefont
			{Kamilla}},\ }\bibfield  {title} {\bibinfo {title} {Quantitative study of
			large composite-fermion systems},\ }\href
	{https://doi.org/10.1103/PhysRevB.55.R4895} {\bibfield  {journal} {\bibinfo
			{journal} {Phys. Rev. B}\ }\textbf {\bibinfo {volume} {55}},\ \bibinfo
		{pages} {R4895} (\bibinfo {year} {1997}{\natexlab{b}})}\BibitemShut {NoStop}%
	\bibitem [{\citenamefont {Chang}\ \emph {et~al.}(2005)\citenamefont {Chang},
		\citenamefont {Regnault}, \citenamefont {Jolicoeur},\ and\ \citenamefont
		{Jain}}]{Chang_PRA_2005}%
	\BibitemOpen
	\bibfield  {author} {\bibinfo {author} {\bibfnamefont {C.-C.}\ \bibnamefont
			{Chang}}, \bibinfo {author} {\bibfnamefont {N.}~\bibnamefont {Regnault}},
		\bibinfo {author} {\bibfnamefont {T.}~\bibnamefont {Jolicoeur}},\ and\
		\bibinfo {author} {\bibfnamefont {J.~K.}\ \bibnamefont {Jain}},\ }\bibfield
	{title} {\bibinfo {title} {Composite fermionization of bosons in rapidly
			rotating atomic traps},\ }\href {https://doi.org/10.1103/PhysRevA.72.013611}
	{\bibfield  {journal} {\bibinfo  {journal} {Phys. Rev. A}\ }\textbf {\bibinfo
			{volume} {72}},\ \bibinfo {pages} {013611} (\bibinfo {year}
		{2005})}\BibitemShut {NoStop}%
	\bibitem [{SM()}]{SM}%
	\BibitemOpen
	\href@noop {} {\bibinfo {title} {See supplemental material at [url will be
			inserted by publisher] for a more comprehensive discussion on the
			quasiparticle spin. supplemental material contains also
			ref~\citenum{Balachandran_1993}}}\BibitemShut {NoStop}%
	\bibitem [{\citenamefont {Hansson}\ \emph {et~al.}(2017)\citenamefont
		{Hansson}, \citenamefont {Hermanns}, \citenamefont {Simon},\ and\
		\citenamefont {Viefers}}]{Hansson_RMP_2017}%
	\BibitemOpen
	\bibfield  {author} {\bibinfo {author} {\bibfnamefont {T.~H.}\ \bibnamefont
			{Hansson}}, \bibinfo {author} {\bibfnamefont {M.}~\bibnamefont {Hermanns}},
		\bibinfo {author} {\bibfnamefont {S.~H.}\ \bibnamefont {Simon}},\ and\
		\bibinfo {author} {\bibfnamefont {S.~F.}\ \bibnamefont {Viefers}},\
	}\bibfield  {title} {\bibinfo {title} {Quantum hall physics: Hierarchies and
			conformal field theory techniques},\ }\href
	{https://doi.org/10.1103/RevModPhys.89.025005} {\bibfield  {journal}
		{\bibinfo  {journal} {Rev. Mod. Phys.}\ }\textbf {\bibinfo {volume} {89}},\
		\bibinfo {pages} {025005} (\bibinfo {year} {2017})}\BibitemShut {NoStop}%
	\bibitem [{\citenamefont {Metropolis}\ and\ \citenamefont
		{Ulam}(1949)}]{Metropolis_1949}%
	\BibitemOpen
	\bibfield  {author} {\bibinfo {author} {\bibfnamefont {N.}~\bibnamefont
			{Metropolis}}\ and\ \bibinfo {author} {\bibfnamefont {S.}~\bibnamefont
			{Ulam}},\ }\bibfield  {title} {\bibinfo {title} {The monte carlo method},\
	}\href {https://doi.org/10.2307/2280232} {\bibfield  {journal} {\bibinfo
			{journal} {Journal of the American statistical association}\ }\textbf
		{\bibinfo {volume} {44}},\ \bibinfo {pages} {335} (\bibinfo {year}
		{1949})}\BibitemShut {NoStop}%
	\bibitem [{\citenamefont {Hastings}(1970)}]{Hastings_1970}%
	\BibitemOpen
	\bibfield  {author} {\bibinfo {author} {\bibfnamefont {W.~K.}\ \bibnamefont
			{Hastings}},\ }\bibfield  {title} {\bibinfo {title} {Monte carlo sampling
			methods using markov chains and their applications},\ }\href
	{https://doi.org/10.2307/2334940} {\bibfield  {journal} {\bibinfo  {journal}
			{Biometrika}\ }\textbf {\bibinfo {volume} {57}},\ \bibinfo {pages} {97}
		(\bibinfo {year} {1970})}\BibitemShut {NoStop}%
	\bibitem [{\citenamefont {Umucalılar}\ and\ \citenamefont
		{Carusotto}(2013)}]{Umucalilar_PhysLettA_2013}%
	\BibitemOpen
	\bibfield  {author} {\bibinfo {author} {\bibfnamefont {R.}~\bibnamefont
			{Umucalılar}}\ and\ \bibinfo {author} {\bibfnamefont {I.}~\bibnamefont
			{Carusotto}},\ }\bibfield  {title} {\bibinfo {title} {Many-body braiding
			phases in a rotating strongly correlated photon gas},\ }\href
	{https://doi.org/https://doi.org/10.1016/j.physleta.2013.06.011} {\bibfield
		{journal} {\bibinfo  {journal} {Physics Letters A}\ }\textbf {\bibinfo
			{volume} {377}},\ \bibinfo {pages} {2074} (\bibinfo {year}
		{2013})}\BibitemShut {NoStop}%
	\bibitem [{\citenamefont {Umucal\ifmmode \imath \else~\i \fi{}lar}\ \emph
		{et~al.}(2018)\citenamefont {Umucal\ifmmode \imath \else~\i \fi{}lar},
		\citenamefont {Macaluso}, \citenamefont {Comparin},\ and\ \citenamefont
		{Carusotto}}]{Umucalilar_PRL_2018}%
	\BibitemOpen
	\bibfield  {author} {\bibinfo {author} {\bibfnamefont {R.~O.}\ \bibnamefont
			{Umucal\ifmmode \imath \else~\i \fi{}lar}}, \bibinfo {author} {\bibfnamefont
			{E.}~\bibnamefont {Macaluso}}, \bibinfo {author} {\bibfnamefont
			{T.}~\bibnamefont {Comparin}},\ and\ \bibinfo {author} {\bibfnamefont
			{I.}~\bibnamefont {Carusotto}},\ }\bibfield  {title} {\bibinfo {title}
		{Time-of-flight measurements as a possible method to observe anyonic
			statistics},\ }\href {https://doi.org/10.1103/PhysRevLett.120.230403}
	{\bibfield  {journal} {\bibinfo  {journal} {Phys. Rev. Lett.}\ }\textbf
		{\bibinfo {volume} {120}},\ \bibinfo {pages} {230403} (\bibinfo {year}
		{2018})}\BibitemShut {NoStop}%
	\bibitem [{\citenamefont {Umucal\ifmmode \imath \else~\i
			\fi{}lar}(2018)}]{Umucalilar_PRA_2018}%
	\BibitemOpen
	\bibfield  {author} {\bibinfo {author} {\bibfnamefont {R.~O.}\ \bibnamefont
			{Umucal\ifmmode \imath \else~\i \fi{}lar}},\ }\bibfield  {title} {\bibinfo
		{title} {Real-space probe for lattice quasiholes},\ }\href
	{https://doi.org/10.1103/PhysRevA.98.063629} {\bibfield  {journal} {\bibinfo
			{journal} {Phys. Rev. A}\ }\textbf {\bibinfo {volume} {98}},\ \bibinfo
		{pages} {063629} (\bibinfo {year} {2018})}\BibitemShut {NoStop}%
	\bibitem [{\citenamefont {Macaluso}\ \emph {et~al.}(2019)\citenamefont
		{Macaluso}, \citenamefont {Comparin}, \citenamefont {Mazza},\ and\
		\citenamefont {Carusotto}}]{Macaluso_PRL_2019}%
	\BibitemOpen
	\bibfield  {author} {\bibinfo {author} {\bibfnamefont {E.}~\bibnamefont
			{Macaluso}}, \bibinfo {author} {\bibfnamefont {T.}~\bibnamefont {Comparin}},
		\bibinfo {author} {\bibfnamefont {L.}~\bibnamefont {Mazza}},\ and\ \bibinfo
		{author} {\bibfnamefont {I.}~\bibnamefont {Carusotto}},\ }\bibfield  {title}
	{\bibinfo {title} {Fusion channels of non-abelian anyons from
			angular-momentum and density-profile measurements},\ }\href
	{https://doi.org/10.1103/PhysRevLett.123.266801} {\bibfield  {journal}
		{\bibinfo  {journal} {Phys. Rev. Lett.}\ }\textbf {\bibinfo {volume} {123}},\
		\bibinfo {pages} {266801} (\bibinfo {year} {2019})}\BibitemShut {NoStop}%
	\bibitem [{\citenamefont {Balachandran}\ \emph {et~al.}(1993)\citenamefont
		{Balachandran}, \citenamefont {Daughton}, \citenamefont {Gu}, \citenamefont
		{Sorkin}, \citenamefont {Marmo},\ and\ \citenamefont
		{Srivastava}}]{Balachandran_1993}%
	\BibitemOpen
	\bibfield  {author} {\bibinfo {author} {\bibfnamefont {A.~P.}\ \bibnamefont
			{Balachandran}}, \bibinfo {author} {\bibfnamefont {A.}~\bibnamefont
			{Daughton}}, \bibinfo {author} {\bibfnamefont {Z.-C.}\ \bibnamefont {Gu}},
		\bibinfo {author} {\bibfnamefont {R.}~\bibnamefont {Sorkin}}, \bibinfo
		{author} {\bibfnamefont {G.}~\bibnamefont {Marmo}},\ and\ \bibinfo {author}
		{\bibfnamefont {A.~M.}\ \bibnamefont {Srivastava}},\ }\bibfield  {title}
	{\bibinfo {title} {Spin--statistics theorems without relativity or field
			theory},\ }\href {https://doi.org/10.1142/S0217751X93001223} {\bibfield
		{journal} {\bibinfo  {journal} {International Journal of Modern Physics A}\
		}\textbf {\bibinfo {volume} {8}},\ \bibinfo {pages} {2993} (\bibinfo {year}
		{1993})}\BibitemShut {NoStop}%
\end{thebibliography}

%
 
\end{document}